\documentclass[11pt]{article}

\usepackage{amsmath}
\usepackage{amssymb}
\usepackage[margin=0.95in]{geometry}
\usepackage{setspace}
\usepackage{natbib}
\usepackage{tikz-cd}
\usepackage[usenames,dvipsnames]{xcolor}
\usepackage{hyperref}
\hypersetup{
    colorlinks=true,
    linkcolor=MidnightBlue,
    citecolor=MidnightBlue,
    urlcolor=MidnightBlue
}

\usepackage{amsthm}
\usepackage{authblk}
\theoremstyle{plain}
\newtheorem{theorem}{Theorem}[section]

\newcommand{\LCDM}{\ensuremath{\Lambda}CDM}
\newcommand{\wCDM}{\ensuremath{w_0 w_a}CDM}
\newcommand{\weff}{\ensuremath{w_{\mathrm{eff}}}}

\title{Does DESI Provide Evidence for Dynamical Dark Energy?}

\author[1,2,3]{Nicola Bamonti \thanks{nicbamonti@gmail.com}}
\affil[1]{Department of Physics, Università di Roma Tor Vergata, Via della Ricerca Scientifica 1, 00133,
Roma, Italy}
\affil[2]{Department of Philosophy, Scuola Normale Superiore, Piazza dei Cavalieri, 7, Pisa, 56126, Italy}
\affil[3]{Department of Philosophy, University of Geneva, 5 rue de Candolle, 1211 Geneva 4, Switzerland}

\date{ }

\begin{document}

\maketitle

\begin{abstract}
\noindent
Recent results from the Dark Energy Spectroscopic Instrument (DESI) have suggested that dark energy, long considered to be a cosmological constant, may actually be \lq{}dynamical\rq{}. To clarify what follows from these results, we distinguish three claims: (D1) the data disfavour the \LCDM\ expansion history; (D2) within the phenomenological Chevallier--Polarski--Linder (CPL) parametrisation, the data prefer $(w_0,w_a)\neq(-1,0)$; and (D3) there exist one or more genuine dynamical dark-energy degrees of freedom in the matter-sector, within GR and coupled to the standard sectors only gravitationally, whose dynamics account for the departure from the \LCDM\ expansion history. 
We demonstrate that DESI establishes D2, which supports D1, but does not establish D3. The \lq{}designer\rq{} construction in $f(R)$ gravity provides a gravitational realisation of the same background history, so that no observable determined solely by that history can discriminate, even in principle, between its \lq{}matter\rq{} and \lq{}gravitational\rq{} readings. Furthermore, non-minimally coupled models admit regular effective phantom-crossing realisations of the relevant phenomenology, while the selected crossing places additional pressure on a single-component, minimally coupled realisation of the matter-sector reading. We characterise the resulting underdetermination between matter and gravitational interpretations---exact at the level of background history, but breakable beyond it---and draw the corresponding norm for survey reporting.
\end{abstract}

\section{Introduction}\label{sec:intro}

For a quarter of a century, the standard model of cosmology has rested on a remarkably simple account of cosmic acceleration: a cosmological constant $\Lambda$, equivalent to a perfect fluid with constant equation of state $w \equiv p/\rho = -1$, supplemented by cold dark matter (CDM) within general relativity (GR). The resulting \LCDM\ model, with $\Omega_m \simeq 0.3$ and $\Omega_\Lambda \simeq 0.7$ in a spatially flat universe ($|\Omega_k| \lesssim 10^{-2}$), has fit essentially all cosmological data since the groundbreaking work of \cite{Riess1998}, the persistent Hubble and $\sigma_8$ tensions notwithstanding \citep{Efstathioutensions,Wolf2026EJPS}.

This consensus has recently been challenged. When the first and second Dark Energy Spectroscopic Instrument (DESI) baryon acoustic oscillation (BAO) data releases \citep{DESI2024,DESI2025} are combined with cosmic microwave background (CMB) data and type\~Ia supernova (SN\~Ia) compilations, the resulting constraints, analysed within the two-parameter Chevallier--Polarski--Linder (CPL) parametrisation \citep{Chevallier2001,Linder2003},
\begin{equation}
w(a)=w_0+w_a(1-a),\label{eq:cpl}
\end{equation}
prefer values away from the \LCDM\ value $(w_0,w_a)=(-1,0)$. For DESI data release 2 (DR2) combined with CMB data, this preference is approximately $3.1\sigma$. Adding SN\~Ia samples gives $2.8\sigma$, $3.8\sigma$, and $4.2\sigma$ for Pantheon+, Union3, and DESY5 respectively, with representative best-fit values $(w_0,w_a)=(-0.752\pm0.057,,-0.86^{+0.23}_{-0.20})$ for the DESY5 combination \citep{Capozziello2026}. Across these combinations, the preferred region lies in the quadrant $w_0>-1$, $w_a<0$, with $w_0+w_a<-1$. The corresponding fitted $w(a)$ is therefore \textit{quintessence-like} today, with $w>-1$, but \emph{phantom} at higher redshift, crossing $w=-1$ at approximately $z_c\approx0.3$--$0.5$. These results have been widely discussed, both in the technical literature and in public communication, as suggesting that dark energy may evolve with time \citep{DESI2025,Castelvecchi2025}. The aim of this paper is to clarify what, if anything, follows ontologically from that description.

Crucially, the DESI technical analysis explicitly states that the CPL parametrisation does not stem from a specific underlying physical model and treats it as a flexible phenomenological description. It also discusses more complex realisations of the preferred phenomenology in the context of dark-energy and modified-gravity \citep{DESI2025}. Our aim is therefore not to attribute D3 below to the DESI collaboration, but to distinguish the parametric result from a stronger ontological interpretation of it. In fact, the CPL functional form itself does not provide that interpretation: the same background phenomenology can result from different underlying physics. Nor do we suggest that the DESI Collaboration regards the present evidence as a final diagnosis: it explicitly anticipates further scrutiny from forthcoming DESI analyses and other experiments, while Euclid will provide complementary measurements of both expansion and structure formation \citep{Castelvecchi2025,Euclid2025}. 
Accordingly, DESI measurements are \textit{not} in question, and \LCDM\ may well not survive them. To distinguish what the measurements establish from stronger interpretations of them, we shall distinguish three claims: 

\begin{description}
\item[(D1)] \emph{Background anomaly}: the data disfavour the \LCDM\ background expansion history $H(z)$, given standard assumptions (FLRW geometry, standard early-universe physics, reliable data). 
\item[(D2)] \emph{Parametric preference}: within the phenomenological \wCDM CPL model class, the data prefer parameter values $(w_0, w_a) \neq (-1, 0)$.
\item[(D3)] \emph{Ontological claim}: there exist one or more genuine dynamical dark-energy \textit{degrees of freedom} in the matter-sector, within GR and coupled to the standard sectors only gravitationally, whose dynamics account for the departure from the $\Lambda CDM$ expansion history. 
\end{description}

What DESI-based cosmological analyses establish is D2 and D2 supports D1 (as we discuss in \S\ref{sec:threeclaims}). But D2---and this is our main claim---does \textit{not} establish D3. The same background expansion history can be generated \textit{exactly} within modified theories of gravity that contain no independent dark energy degree of freedom whatsoever in the matter-sector, and the same fitted evolution can be read as an effective representation of modified gravitational dynamics in fluid vocabulary. Accordingly, if the `evidence for dynamical dark energy' formulation is read ontologically as D3, it is elliptical at best: what the background data provide is evidence of a departure from the \LCDM\ \emph{expansion history}, while the physical interpretation---new matter-sector dynamics or modified gravitational dynamics---is not something the background channel could ever settle. This does not imply empirical equivalence \textit{simpliciter}. The full CMB data already contain perturbation-sensitive information, and further non-background observables can in principle discriminate between specified rivals \citep{DESIMG2024,DESI2025}. Our claim is instead that the underdetermination is exact for the background channel, while the presently available discriminating perturbative evidence has not eliminated the relevant gravitational alternatives (\S\ref{sec:breaking}).

The fact that \lq{}modified gravity can mimic dark energy\rq{} is not news to physicists, who have established it with explicit constructions \citep{NOJIRI2007,Joyce2016}.  One such result goes beyond mimicry. The \lq{}designer\rq{} reconstruction programme in $f(R)$ gravity shows that, for a target expansion history $H(z)$, one can construct gravitational theories that reproduce the same background \textit{exactly} \citep{Song2007,Pogosian2008,Sotiriou2010}. We recast this construction as a degeneracy theorem concerning the evidential scope of geometric background data. In other words, no observable fixed solely by the  background history can distinguish between matter-sector and gravitational alternatives.\footnote{Here, we use 'observable' in its most deflated operational sense. While the term is conceptually loaded in the foundations of physics, our analysis is independent of discussions concerning observability in GR. For a philosophical analysis of the term in the context of GR and modern cosmology, see, e.g., \cite{PITTS20221,BamontiThebault2024} and references  therein.}

The gravitational reading is not merely a formal possibility. Recent work has shown that non-minimally coupled scalar-tensor theories---which fall on the gravitational side of the distinction adopted here---can accommodate the relevant phenomenology, including phantom crossing, and fit the relevant DESI combinations comparably to the \wCDM\ template \citep{Ye2025,Wolf2025}. The designer construction and these concrete models therefore play complementary roles: the former establishes the exact background degeneracy, while the latter shows that the gravitational side of the rivalry contains phenomenologically competitive candidates.

We also use the phantom crossing as a test of a possible theoretical preference for the matter-sector reading (\S\ref{sec:phantom}). If, having recognised the empirical underdetermination, one were to privilege the matter-sector reading on the basis that it offers the simplest and most direct theoretical realisation of the reconstructed phenomenology, the phantom crossing undermines that rationale by showing that this is not possible using its simplest single-component, minimally coupled realisation.  Therefore, reproducing the crossing within D3 requires additional structure. Gravitational-sector models, by contrast, admit stable effective-crossing realisations without encountering the difficulties that affect the simplest single-field matter-sector realisations. This does not establish a general preference for modified gravity over D3, but it removes one possible reason for favouring the simplest matter-sector realisation. 

We then situate the case within recent work on underdetermination in cosmology \citep{FerreiraWolfRead2025,WolfRead2026}. 
That literature has emphasised the difficulty of inferring underlying microphysics from a restricted set of cosmological observables. This includes cases in which distinct models within a shared framework become arbitrarily close empirically.\footnote{See also \citet{WolfDuerr2024} who locate the problem within the competing approaches to cosmic acceleration, including modified gravity, but ask which of them merits further investigation, rather than what the evidence can settle.} The DESI case presented here has a different structure: the rivalry crosses the \lq{}matter--gravity divide\rq{}, and the degeneracy is exact for the background history while remaining breakable in principle through sufficiently discriminating non-background probes. Section~\ref{sec:philosophy} develops this comparison in detail.

Finally, we distinguish the isolated interpretive problem from the robustness of the anomaly itself. If the departure from \LCDM\ weakens, there is no explanandum to explain. However, if it is strengthened by future data, the background-level underdetermination nevertheless remains. 
Therefore the robustness of the anomaly and the identification of its physical source are separate questions. This separation motivates the introduction of modest reporting norm defended below: phenomenological constraints on the expansion history should be distinguished from stronger claims about what physically produces that history. The latter should be based on evidence from channels capable of discriminating between the rival dynamics, rather than on increased precision within the shared background description.

The paper proceeds as follows. Section~\ref{sec:desi} reconstructs the inferential structure of the DESI result, from raw observables through the CPL parametrisation to its interpretation as evidence for evolving dark energy, and identifies precisely where theoretical assumptions enter. Section~\ref{sec:mg} develops the modified-gravity alternative: the effective-fluid representation of $f(R)$ gravity, the designer degeneracy theorem, and an assessment of the viability constraints on concrete models. Section~\ref{sec:phantom} presents the phantom-crossing argument. Section~\ref{sec:philosophy} turns to the epistemology of the case: why the reporting convention invites the slide, what kind of underdetermination results, how it can be broken, and how secure the anomaly under debate really is. 

\section{The inferential structure of the DESI result}\label{sec:desi}

\subsection{From BAO observables to the expansion history}\label{sec:observables}

It is worth specifying what DESI measures, because interpreting the dynamical dark energy result involves several inferential steps, and the argument below relies on observables which are determined by the background expansion history. 

DESI is optimised to measure the BAO feature by constraining the apparent angular and radial extent of a characteristic comoving scale imprinted on the clustering of matter: the sound horizon at the baryon-drag epoch, $r_d$. Concretely, at a given redshift, the survey constrains dimensionless ratios such as $D(z)/r_d$ and $D_H(z)/r_d$, where, in spatially flat universe, $D(z) = \int_0^z c\,dz'/H(z')$ is the transverse comoving distance and $D_H(z) = c/H(z)$ is the Hubble distance. The scale $r_d$ itself is not \textit{directly} determined by BAO data alone. In practice, calibrating $r_d$ requires assumptions and information about early-universe physics, typically supplied by CMB constraints on the physical baryon and matter densities, or by BBN information in combination with the cosmological data.. Crucially, at the level of the reported BAO distance likelihood, BAO observables are \textit{geometric} in nature: they are sensitive only to the background expansion history $H(z)$ (and to the calibration scale $r_d$, which is supplied by early-universe physics). The data constrain geometric quantities, such as distances, expansion rates, and angular scales, not the properties of a dark-energy \lq{}fluid\rq{}.\footnote{Strictly speaking, the \emph{extraction} of the BAO scale from galaxy  clustering involves perturbative physics, such as nonlinear evolution of the acoustic peak and density-field reconstruction techniques, that presuppose GR-like growth. This does not threaten the argument. The induced scale shifts are sub-percent, and the boundary condition of GR recovery at early times leaves $r_d$ untouched, so BAO function as geometric probes to the accuracy at which DESI reports them. Moreover, the geometric idealisation is derived from the survey itself: the background constraints are based on the distance likelihood under fixed assumptions about the perturbative sector. Therefore, any perturbative information used in the BAO extraction does not provide an independent means of distinguishing between the matter-sector and gravitational interpretations considered here.}

In fact, this list does not include any dark-energy equations of state. The quantity $w(z)$ is not measured, it is \emph{reconstructed}, and the reconstruction is doubly model-dependent. The contrast we intend is not the naive one between theory-free and theory-laden quantities---every quantity in this chain, including $D/r_d$ and $H(z)$ itself, is model-mediated, and we return to the point in \S\ref{sec:theoryladen}. The operative distinction is \emph{invariance across the rival frameworks in play}: $H(z)$ is common ground between the matter and gravitational readings, both of which are FLRW cosmologies, whereas interpreting $w_X(z)$ as the equation of state of genuine matter-sector degrees of freedom requires the former. It is in this framework-relative sense that $w(z)$ is reconstructed rather than measured. 

Assume GR, spatial flatness, standard pressureless matter-sector $(w_m=0)$, and an additional smooth dark-energy sector $X$ not interacting with matter and represented at the background level as an effective fluid. The Friedmann equation reads
\begin{equation}
H^2(z) = H_0^2 \left[ \Omega_m (1+z)^3 + \Omega_X \frac{\rho_X(z)}{\rho_{X,0}} \right],
\label{eq:friedmann}
\end{equation}
and the continuity equation for the $X$- component 
\begin{equation}\label{eqstateX}
\dot\rho_X + 3H(1+w_X)\rho_X = 0
\end{equation}
gives
\begin{equation}
\rho_X(z) = \rho_{X,0} \exp\!\left[ 3 \int_0^z \frac{1 + w_X(z')}{1+z'}\, dz' \right].
\label{eq:rhoDE}
\end{equation}
A constant energy density requires $w_X = -1$ exactly; accelerated expansion requires only $w_{\mathrm{tot}} < -1/3$. Combining eqs. \eqref{eq:friedmann}
 and \eqref{eqstateX}, and assuming a separately conserved matter component with $\rho_m \propto (1+z)^3$, one may formally reconstruct the equation of state from the expansion history (cf. \citealp[eq.7]{SahniStarobinsky2006}):
\begin{equation}
w_X(z) = \frac{\tfrac{2}{3}(1+z)\, H H' - H^2}{H^2 - H_0^2\, \Omega_m (1+z)^3},
\label{eq:reconstruction}
\end{equation}
where $H' = dH/dz$. Equation~\eqref{eq:reconstruction} makes three things explicit. First, $w_X(z)$ involves a \emph{derivative} of $H(z)$. Reconstructing $w_X(z)$ requires differentiating noisy and sparsely sampled data, making the problem intrinsically ill-conditioned. 
This is precisely why parametrisations such as CPL are used in practice. Second, the reconstruction is degenerate with $\Omega_m$: a shift in the assumed matter density propagates directly into the inferred $w(z)$, and indeed part of the DESI preference for the \wCDM model may reflect a mismatch between the $\Omega_m$ values favoured by different cosmological probes under \LCDM\ \citep{DESI2024}.\footnote{The geometric degeneracy between the matter density and the dark energy parameters is a cornerstone of observational cosmology. See, e.g., \cite{Efstathiou1999,Serjeant2010-ue,PLANCK2020}.} Third, and most importantly for what follows, the left-hand side of \eqref{eq:reconstruction} is the equation of state \emph{of whatever the model assigns to the right-hand side}. The reconstruction therefore inherits all the assumptions built into that decomposition.\footnote{If that sector contains several genuine matter-sector degrees of freedom, $w_X$ may characterise their aggregate background stress-energy without being the microphysical equation of state of any single constituent.} If the true dynamics is not captured by this GR-plus-matter-sector decomposition---in particular, if the same background evolution is generated by modified gravitational dynamics---equation~\eqref{eq:reconstruction} \textit{still} returns a function $w_X(z)$, but that function then need not correspond to the pressure-to-density ratio of any genuine matter-sector source. Rather, it is an effective quantity that parametrises how the underlying dynamics would appear when projected onto a dark-energy fluid template. 

\subsection{The CPL parametrisation as an instrument}
\label{sec:cpl}

The CPL parametrisation \eqref{eq:cpl} is a numerically convenient, two-parameter phenomenological description of dark-energy evolution: well-behaved at high redshift (where $w \to w_0 + w_a$), linear in the scale factor around the present ($w \to w_0$ at $z = 0$), and capable of approximating the background behaviour of broad classes of scalar-field models \citep{Chevallier2001}. It is not derived from a Lagrangian, a symmetry principle, or an effective-field-theory expansion. It is a fitting template, and the interpretation of $(w_0, w_a)$ as parameters of the microphysical equation of state of a particular matter-sector component is an additional physical interpretation, not part of the measurement. Inserting \eqref{eq:cpl} into \eqref{eq:rhoDE} yields the closed form
\begin{equation}
\rho_X(z) = \rho_{X,0}\, (1+z)^{3(1+w_0+w_a)} \exp\!\left( -\frac{3 w_a z}{1+z} \right),
\label{eq:cplrho}
\end{equation}
which defines the \wCDM model whose two extra parameters are then fit jointly with the standard ones.

It is important to understand what a statement like `$4.2\sigma$ preference for dynamical dark energy' actually means. It is a model-comparison statement within a hypothesis space: \LCDM\ against its two-parameter extension \wCDM, in which it is nested as the point $(w_0,w_a)=(-1,0)$. However, it does not quantify the probability of D3---that the departure is physically due to genuine dynamical matter-sector degrees of freedom---for at least two reasons. 
Firstly, an absolute posterior probability for D3 would require a specified hypothesis space and priors spanning D3 and the relevant alternatives---including modified gravity, interacting dark-sectors, non-trivial spatial geometry, and possible systematics---which the quoted 4.2$\sigma$ comparison does not provide. 
Secondly, the comparison is conditional on the chosen parametrisation and, in Bayesian model selection, on the adopted priors. \cite{CortesLiddle2024}, for instance, argue that plausible changes to the broad flat prior on $(w_0, w_a)$ would reduce the discrepancy with $\Lambda CDM$.\footnote{This conditionality should not be confused with a demonstrated fragility of the DESI trend: the DR2 supporting analysis finds broadly consistent behaviour across several alternative parametrisations and non-parametric reconstructions. For further discussion on this prior-based deflation, see \cite{Shlivko2024}.} 
The number `$4.2\sigma$' is therefore a statement about the relative fit of \LCDM\ and \wCDM\ in this nested comparison---a fact about D2, not a posterior probability for D3.

To avoid any misunderstanding, we point out that restriction to a limited hypothesis space is a generic feature of model comparison and not a defect peculiar to DESI. By itself, it licenses no naive scepticism about ontological conclusions---otherwise the argument would prove too much and invalidate every claim of entity discovery in physics. The observation made here is deliberately modest: the $\sigma$-level does not \emph{encode} the probability of D3.
Furthermore, the reason why in \emph{this} case no background significance level, however high, could establish D3 is not the restriction of the hypothesis space as such. Rather, it is because among the omitted alternatives there exists a class of modified-gravity rivals that is \emph{provably indistinguishable at the level of the background observables}---a burden discharged by the degeneracy theorem of \S\ref{sec:designer}, to which the present section is preparatory. 

\subsection{Three claims and the interpretive step}\label{sec:threeclaims}

We can now state the central distinction. At the level of background history inference, the chain relevant to interpreting the DESI result is as follows:

\begin{equation*}
\begin{array}{@{}r@{}c@{}l@{}}

\text{distances}
\xrightarrow{\;\text{FLRW},\,r_d\;}
H(z)
\xrightarrow{\;\text{Friedmann},\,\Omega_m\;}
\rho_X^{\rm eff}(z)
\xrightarrow{\;\text{continuity}\;}
w_X^{\rm eff}(z)
\xrightarrow{\;\text{CPL}\;}
&
(w_0,w_a)\neq(-1,0)
&

\\[0.35ex]

&
\Updownarrow
&

\\[0.35ex]

\text{(D1)}\;
(\text{non-}\Lambda\mathrm{CDM}\text{ background})
\xleftarrow{\;\text{supports}\;}
&
\text{(D2)}
&

\\[0.8ex]

\text{(D3)}\;
(\text{GR + genuine matter-sector dark energy})
\overset{
  \substack{
    \text{ontological}\\[-0.3ex]
    \text{interpretation}
  }
}{\dashleftarrow}
&
\end{array}
\end{equation*}

Each arrow imports assumptions, but not of the same kind. The first imports background geometry and calibration (FLRW, $r_d$), which is standard, explicitly acknowledged, and common ground between the rival readings at issue. The second and third introduce the GR-plus-effective-fluid decomposition: they define $\rho_X$ as the residual contribution in the Friedmann equation and $w_X$ through the continuity equation. As definitions they provide an innocuous bookkeeping device for the class of theories considered below, but they describe genuine matter-sector stress-energy only if the residual corresponds to one or more genuine matter-sector degrees of freedom (\S\ref{sec:observables}). The fourth arrow imports the adequacy of the CPL compression, which is examined in \S\ref{sec:cpl}. Our concern is the dashed arrow: the transition from a parametric preference (D2) to an ontological claim (D3).\footnote{The chain is a rational reconstruction of the inference, not of the computational pipeline: in practice the CPL form is imposed at the model-specification stage and the $(w_0,w_a)$ posteriors are obtained by fitting the resulting \wCDM\ model directly to the distance data (\S\ref{sec:observables}). The displayed assumptions are thereby imposed jointly rather than traversed sequentially; nothing in what follows depends on the order. Also, the diagram isolates the background-level inferential channel. The full DESI data also contain perturbative information. This does not alter the distinction between D2 and D3, but it means that the exact degeneracy established below applies specifically to the background observables.} This distinction does not attribute D3 to the DESI technical analysis, which explicitly treats CPL as a flexible phenomenological parametrisation rather than as a specific underlying physical model; rather, it asks whether D3 is warranted as an ontological interpretation of D2. 
This transition is \emph{evidentially} licensed by the background data only if the \lq{}material interpretation\rq{} is independently secure---if, that is, the same apparent evolution could not equally be generated by modified gravitational dynamics without independent dynamical dark-energy degrees of freedom in the matter-sector. 
The next section shows that this is possible exactly at the background level through a family of designer gravitational theories. 

Before that, a remark on D1-D2-D3. \LCDM\ sits inside the CPL family as the point $(w_0,w_a)=(-1,0)$, and a preference for other values indicates that the \LCDM\ expansion history accommodates the combined data worse than some alternative history does. D2 then supports D1. However, whether D1 is robust remains unsettled rather than established: its statistical strength depends on the dataset combination, and possible dataset systematics remain under active scrutiny (\S\ref{sec:robustness}). 

The logical structure is thus as follows: a strong ontological interpretation (D3) that does not follow from a genuine but framework-internal parametric result (D2), which in turn supports a weaker empirical assertion (D1).

\section{Modified gravity and the effective dark-energy fluid}\label{sec:mg}

\subsection{$f(R)$ gravity and its fluid representation}
\label{sec:fR}

The cleanest laboratory for the inter-theoretic underdetermination between dynamical-dark-energy and modified-gravity explanations is metric $f(R)$ gravity, the minimal extension of GR in which the Einstein--Hilbert Lagrangian density $R$ is replaced by a function of the Ricci scalar \citep{Buchdahl1970,Sotiriou2010,Nojiri2017}:\footnote{For a philosophically oriented overview, see \cite{DUERR202110}.}
\begin{equation}
S = \int d^4x \sqrt{-g}\, \left[ \frac{f(R)}{2\kappa^2} + \mathcal{L}_m \right], \qquad \kappa^2 = 8\pi G.
\label{eq:fRaction}
\end{equation}
In the Jordan-frame representation, matter couples minimally and the modification is purely gravitational.\footnote{Crucially, $f(R)$ admits an Einstein-frame presentation as GR plus a scalar field and the Einstein-frame scalar is directly coupled to matter. We take up the point in full in \S\ref{sec:taxonomy}.} Variation with respect to the metric yields the fourth-order modified Einstein field equations which, on a flat FLRW background with pressureless matter, reduce to the two modified Friedmann equations \citep{DeFelice2010,Amendola:2015ksp}:\footnote{We omit radiation here to keep the construction transparent. It can be restored without changing the second-order designer structure .}
\begin{align}
3 F H^2 &= \kappa^2 \rho_m + \frac{FR - f}{2} - 3 H \dot F,
\label{eq:fRfriedmann}\\
-2 F \dot H &= \kappa^2 \rho_m + \ddot F - H \dot F,
\label{eq:fRacceleration}
\end{align}
where $F \equiv \partial f/\partial R$ , $R = 6(\dot H + 2H^2)$ and where overdots denote derivatives with respect to cosmic time. The non-linear theory propagates, in addition to the graviton, a single scalar degree of freedom (the `scalaron' $F$), and standard health requirements include $F > 0$ and $f_{RR} \equiv \partial^2 f/\partial R^2 > 0$, ensuring a positive effective gravitational coupling (no ghosts) and avoiding the Dolgov--Kawasaki tachyonic instability, respectively \citep{Amendola2007}. 

The same background equations can now be rewritten in the standard GR form. To do so, the non-Einsteinian curvature terms are moved to the right-hand side and represented as an effective fluid. At the homogeneous level, this is equivalent to defining an effective energy density and pressure by
\begin{align}
\kappa^2 \rho_{\mathrm{eff}} &\equiv \frac{FR - f}{2} - 3 H \dot F + 3 H^2 (1 - F), \label{eq:rhoeff}\\
\kappa^2 p_{\mathrm{eff}} &\equiv \ddot F + 2 H \dot F - \frac{FR - f}{2} - (2\dot H + 3 H^2)(1 - F). \label{eq:peff}
\end{align}
With these bookkeping definitions, equations \eqref{eq:fRfriedmann}--\eqref{eq:fRacceleration} take \emph{exactly} the form of the GR Friedmann equations with an extra fluid: $3H^2 = \kappa^2(\rho_m + \rho_{\mathrm{eff}})$ and $-2\dot H = \kappa^2(\rho_m + \rho_{\mathrm{eff}} + p_{\mathrm{eff}})$, with $\rho_{\mathrm{eff}}$ obeying the standard continuity equation. By virtue of the Bianchi identity and matter conservation, the effective fluid defined above satisfies
\begin{equation}
\dot\rho_{\rm eff}
+
3H(\rho_{\rm eff}+p_{\rm eff})
=0.
\end{equation}
One may then define the effective equation of state
\begin{equation}
w_{\rm eff}(z) \equiv \frac{p_{\mathrm{eff}}}{\rho_{\mathrm{eff}}},
\label{eq:weff}
\end{equation}
which is, in general, a non-trivial function of redshift. 
If the background expansion generated by eq. \eqref{eq:fRaction} is analysed within a GR-plus-fluid framework, the departure from GR can be represented by a non-trivial effective equation of state \weff. When that history is well captured by the CPL template, this can yield D2, but what does not thereby follow is D3. In the Jordan-frame representation there is no additional dark-energy component in the matter-sector: \weff\ parametrises the effective curvature fluid introduced by the rearrangement above, rather than the microphysical equation of state of any genuine matter-sector source.\footnote{The point generalises at the background level well beyond $f(R)$: scalar-tensor theories, braneworld models, and the broader Horndeski class all admit effective-fluid representations with evolving \weff\ \citep{Clifton2012}. We focus on $f(R)$ because it makes the degeneracy theorem of \S\ref{sec:designer} exact and elementary.}

\subsection{The designer construction: a degeneracy theorem}\label{sec:designer}

So far we showed that $f(R)$ modified gravity produces \emph{some} evolving \weff. A stronger claim is that it can be \textit{designed} to reproduce a broad class of evolving \weff, including CPL expansion history inferred from the DESI data combinations. This is the broad content of what is known as the the `designer' $f(R)$ construction programme, which we now present in a form adapted to our purposes \citep{Song2007,Pogosian2008}. 

It is convenient, and standard in designer $f(R)$ reconstructions, to use the e-folding variable $N\equiv\ln a$, for which $d/dt=H\,d/dN$. This is merely a reparametrisation of the background evolution, not an additional physical assumption. 
Fix a target expansion history $H(N)$: for instance, a CPL history from the DESI analyses. We may regard $f$ as a function of $N$, with $F = f_{,N}/R_{,N}$.\footnote{To further clarify: $f(N)$ is a shorthand for the composite function $f(R(N))$.} Along that background, $R(N)$ is known and, away from stationary points where $R_{,N}=0$, the chain rule gives 
\begin{equation}
F=df/dR=f_{,N}/R_{,N}.\label{FN}
\end{equation}
Using $\dot F = H F_{,N}$, the modified first Friedmann equation
\eqref{eq:fRfriedmann} becomes
\begin{equation}
3H^2(F_{,N}+F)+\frac{f-FR}{2}
=
\kappa^2\rho_m.
\end{equation}
Substituting relations \eqref{FN} yields
\begin{equation}
3 H^2 \left( \frac{f_{,N}}{R_{,N}} \right)_{,N} + \left( 3 H^2 - \frac{R}{2} \right) \frac{f_{,N}}{R_{,N}} + \frac{f}{2} = \kappa^2 \rho_m(N),
\label{eq:designer}
\end{equation}
a linear, second-order ordinary differential equation for $f(N)$, whose coefficients are fixed once the target background is specified. Because the equation is second order, the target expansion history does not uniquely determine the gravitational Lagrangian: its solutions differ by integration constants. Hence, wherever the reconstruction is regular, specifying $H(N)$ does not uniquely fix the gravitational dynamics: it determines a family of $f(R)$ theories reproducing the same background expansion.\footnote{The standard designer construction supplements this equation with a high-curvature boundary condition that recovers GR at early times, leaving a family of modified-gravity models with the prescribed background history \citep{Song2007,Pogosian2008}.}

The designer construction has a direct consequence for the present argument, which we formulate as follows

\begin{theorem}
\textbf{Designer theorem (Background degeneracy).} For the class of smooth expansion histories for which the standard designer reconstruction is regular—including the CPL histories considered here—one can construct $f(R)$ models 
that reproduce the prescribed $H(z)$ \emph{exactly} while attributing the departure from \LCDM\ to modified gravitational dynamics rather than to an additional matter-sector dark-energy source.\footnote{As usual in designer reconstructions, this statement is restricted to intervals on which the reconstruction is regular and to backgrounds for which the standard high-curvature GR boundary condition can be imposed \citep{Pogosian2008}.}
Consequently, given the same early-time physics and sound-horizon calibration, no observable determined solely by the homogeneous expansion history---such as BAO distance ratios, SN~Ia distances,\footnote{SN~Ia distances qualify as purely geometric only under the additional assumption that the standardised intrinsic luminosity is unaffected by the gravitational modification \citep{Wright2018}.} or geometrical CMB distance priors---can discriminate, even with arbitrarily precise data, between (i) GR plus a genuine dynamical dark-energy matter-sector with equation of state $w(z)$ and (ii) a designer $f(R)$ theory mimicking the same $H(z)$.
\end{theorem}

The degeneracy is exact and it is confined to the \textit{background channel}: a preference for $w(z)\neq-1$ inferred solely from the expansion history cannot, by itself, identify a genuine, new dark-energy source in the matter-sector as the cause of the deviation. The theorem establishes mathematical \emph{existence}, not  \lq{}physical health\rq{} or \lq{}observational viability\rq{}. A reconstructed theory must still satisfy the conditions required to avoid pathological dynamics, and it must survive tests that probe more than the homogeneous expansion. This does not undermine the designer result, because the degeneracy theorem does not claim that every reconstruction is healthy or observationally viable. Its role is to determine what the background data can discriminate, not which theory deserves belief. 

An attempt to sustain D3 by eliminating the modified-gravity alternative therefore requires two further steps: the alternative must first be physically healthy because the collapse of an ill-posed theory would be no news, and \textit{all} healthy gravitational alternatives reproducing the relevant DESI phenomenology must then be excluded on independent observational grounds. 

\subsubsection{Health, viability, and the scope of the degeneracy theorem}\label{sec:viability}

A natural objection is that the best-known modified-gravity models are already tightly constrained by local and astrophysical tests, and that precisely those constraints prevent them from reproducing the DESI-favoured expansion history. The designer theorem would then establish only a mathematical possibility, with physically relevant alternative collapsing back onto \LCDM.
While this objection points to a genuine constraint, assessing it requires keeping two conditions apart. 

First, a reconstructed theory must be \textit{physically healthy}. In metric $f(R)$ gravity, for instance, standard requirements include $F>0$ and $f_{RR}\equiv\partial^2 f/\partial R^2>0$ introduced above. A pathological reconstruction does not constitute a physically admissible rival. 

Secondly, even a healthy theory may fail \emph{observational viability}. Here, we call a theory viable if the available evidence does not exclude it, including evidence from local and astrophysical tests of gravity.

The distinction matters because a strictly eliminative route to D3 ultimately turns on the latter. Suppose that \textit{every}  healthy gravitational-sector alternative capable of reproducing the DESI-favoured phenomenology were ruled out by observational tests. The gravitational reading would then be off the table, and D3 could be approached by elimination, on a warrant supplied by an independent experimental probe. 
A caveat on \textit{every}: if read as a demand for one test per model, the supposition could never be met, since new models can always be written down. But it is not that demand. It states a condition under which the eliminative route would go through, and a condition of that form carries no implication that anyone could check it case by case. The relevant constraints do not attach to models one at a time either. A Solar-System-bound constraint on fifth forces restricts a coupling strength, and applies to  any theory with that coupling, including theories that have not yet been proposed. Therefore, warrant for the universal premise would have to come from constraints or no-go results that demonstrably cover an exhaustively specified class of gravitational alternatives, rather than from an enumeration of individual models.\footnote{The burden of establishing such universal exclusion lies with the eliminative route. For our negative conclusion, it is sufficient that this burden has not been met; the existence of even a single healthy rival would provide a stronger counterexample.}

Also consider that showing that a particular reconstruction is unhealthy, or that a particular healthy model fails observational tests, establishes only the failure of \textit{that} model. What matters is thus whether healthy gravitational alternatives to the relevant cosmological deviation remain observationally viable. 

One of the flagship $f(R)$ examples may help illustrate the point. 
Among the best-known viable $f(R)$ models is that of \cite{HuSawicki2007},\footnote{$f(R) = R - m^2\, c_1
(R/m^2)^n / \left[1 + c_2 (R/m^2)^n\right]$, with $m^2 \equiv
\kappa^2 \bar\rho_{m,0}/3$. Another well-known model is Starobinsky's disappearing-cosmological-constant model \citep{Starobinsky2007}.} which is designed to recover GR at high curvature while generating late-time acceleration, and to pass Solar System tests through the chameleon screening mechanism \citep{Khoury2004}. However, screening strongly restricts the allowed departure from GR: remaining compatible with Galactic and Solar-System constraints pushes the field amplitude to roughly $|f_{R0}|\lesssim10^{-6}$, in a regime in which the background expansion is extremely close to \LCDM. Thus the screened Hu–Sawicki model cannot realise a background deviation of the DESI amplitude. 

Conceding this does \textit{not} rescue the inference to D3. It shows that Hu--Sawicki is not the required rival: in the regime in which it survives local constraints, it does not generate the cosmological departure at issue. But excluding this particular model does not exclude the modified-gravity reading as such. 
Conversely, blocking that route requires existence: one healthy gravitational-sector account of the DESI-favoured phenomenology must exist.\footnote{Exhibiting one does not show D3 to be false. What the exhibit removes is the warrant that the eliminative route would supply, and it is the warrant that is at issue here.} That the Hu--Sawicki $f(R)$ model is not such an account leaves the requirement untouched. The role of Hu--Sawicki in our argument is thus methodological rather than phenomenological. 

There are, moreover, gravitational models that come closer to meeting this requirement. Within the $F(R)$ family, recent work has exhibited models satisfying the viability conditions imposed in the analysis and providing competitive fits to DESI and other late-time data while producing an effective dark-energy evolution qualitatively similar to the phenomenology at issue \citep{Odintsov2025,Odintsov2026}. More generally, non-minimally coupled gravity has been shown to provide comparably good fits to the DESI data combinations, with non-minimal coupling improving the fit relative to minimally coupled alternatives in the models considered \citep{Ye2025,Wolf2025}.\footnote{Nor does the post-DESI provenance of those fits disqualify them as candidates: the fluid reading is symmetrically placed, with the CPL template that predates the anomaly and concrete realisations of the preferred $w(z)$ built after it \citep{Cai2026}. On both readings, a pre-existing theory class is fitted to the data once the data arrive; post-hoc realisation is the common condition of the rivalry, not a liability of one side.}
These models face a serious local-test problem: a coupling of the strength required by the cosmological fit would generate fifth forces not observed in the Solar System, so a screening mechanism is required. Whether such a mechanism can be implemented satisfactorily remains an open model-building question.\footnote{\citeauthor{Ye2025} (\citeyear{Ye2025}) reconstruct the equation of state and the gravitational-sector without imposing a template, and interpret the result as a hint of modified gravity, assuming the modification to be screened on small scales. \citeauthor{Wolf2025} (\citeyear{Wolf2025}) find that the coupling preferred cosmologically exceeds the Cassini/MESSENGER bound by roughly four orders of magnitude. They show that Solar-System screening can in principle be obtained by supplementing the fitted action with a higher-derivative operator generating Vainshtein screening and defer the detailed construction to further work.}
Their observational viability is therefore unsettled. But unsettled is not excluded, and the pressure they face comes precisely from the independent local channel that \S\ref{sec:breaking} identifies as one of the routes along which the background degeneracy is broken. A strictly eliminative route to D3 is therefore not supplied by the background evidence alone. 

\section{The phantom divide as a stress test of the matter-sector interpretation}\label{sec:phantom}

The designer theorem establishes that the background evidence does not discriminate between the matter-sector and gravitational readings. One might nevertheless regard the former as the theoretically more straightforward interpretation, and modified gravity as merely a formally available alternative. The phantom crossing provides a useful test of that presumption. It does not discriminate modified gravity from D3 as a whole,  rather it shows that the most immediate, single-component realisation of the matter-sector reading is itself theoretically non-trivial.

The DESI-preferred parameter region has a sharp qualitative signature. With $w_0 > -1$ and $w_0 + w_a < -1$, the reconstructed $w(z)$ is phantom at redshifts above $z_c$ and quintessence-like below, crossing the \emph{phantom divide} $w = -1$ at $z_c \approx 0.3$--$0.5$ depending on the data combination (see \citealp[\S 5]{Capozziello2026} and references therein). Non-parametric reconstructions show a similar preference for phantom-crossing behaviour, although they do not establish that the crossing is genuine rather than an effective feature of the reconstruction \citep{Lodha2025}. Taking a \textit{single-component}, minimally coupled fluid realisation at face value, the preferred reconstructed $w(z)$ would correspond to a component that violated the null energy condition ($\rho+p<0$) in the recent past and has since ceased to do so.

We argue that the phantom demand within the GR-plus-single-component framework is costly. For a single canonical scalar field minimally coupled to gravity---quintessence, the paradigm of dynamical dark energy \citep{Tsujikawa2013}---one has
\begin{equation}
\rho_\phi + p_\phi = \dot\phi^2 \geq 0 \quad \Longrightarrow \quad w_\phi \geq -1,
\end{equation}
the crossing is \emph{kinematically impossible}, whatever the potential. The result extends beyond quintessence: for a single perfect fluid, or a single k-essence field, the crossing point $w = -1$ is a singular point of the perturbation dynamics---the standard fluid/field perturbation equations become pathological there, with divergent or unstable behaviour. Thus, a regular crossing for the matter sector is possible in general, but requires additional structure---for example multiple degrees of freedom, higher-derivative terms, interactions, or departure from the minimally coupled single-component framework \citep{Vikman2005,Kunz2006}. 
This is why, in the wake of the DESI releases, the model-building literature has explored a range of more elaborate realisations \citep{Cai2026}. The point is therefore narrower: the phantom crossing places particular pressure on the simplest single-component, minimally coupled realisation of D3, not on D3 as such.

In the $f(R)$ representation developed above, the quantity that `crosses' is the derived effective ratio \eqref{eq:weff}, not the microphysical equation of state of a matter-sector degree of freedom. The crossing condition $\rho_{\mathrm{eff}} + p_{\mathrm{eff}} = 0$ reads, from \eqref{eq:rhoeff}--\eqref{eq:peff},
\begin{equation}
\ddot F - H \dot F - 2 \dot H\, (1 - F) = 0,
\end{equation}
which is a regular condition on the background evolution. In a healthy $f(R)$ model, crossing of $w_{\rm eff}$ does not by itself signal a divergence, a ghost, or energy condition violation by a matter-sector field.\footnote{As we will stress in \S\ref{sec:taxonomy}, the point is independent of the choice between Einstein and Jordan frame.} 
Indeed, it is a classic result that $f(R)$ models of late-time acceleration \emph{generically} display phantom behavior of \weff\ followed by a crossing of the divide. 
This possibility was established well before DESI \citep{Amendola2008}. 
Post-DESI analyses reinforce the broader point: stable realisations of the relevant crossing can be obtained in extended gravitational frameworks, including non-minimally coupled gravity \citep{Ye2025,YeCai2025}. Most recently, \cite{GarciaGarcia26} found that their non-minimally coupled scalar-field models that permit phantom crossing fit the data better than minimally coupled quintessence. 

\subsection{A test of theoretical straightforwardness}
\label{sec:asymmetry}

We can now assemble the pieces into an argument:
\begin{enumerate}
\item The DESI-preferred reconstructions exhibit a phantom-crossing feature at the level of the effective background equation of state.
\item If this feature is interpreted within D3, canonical quintessence---the simplest single-component, minimally coupled realisation---cannot reproduce it, and broad classes of single-field models encounter pathologies at the crossing.
\item In the gravitational sector, by contrast, there exist realisations in which the corresponding effective crossing can occur without the pathologies encountered by the simplest matter-sector realisations. 
\item Hence, if theoretical simplicity is invoked to privilege the matter-sector reading, the phantom feature undermines that rationale: reproducing the crossing within D3 requires moving beyond its simplest implementation, whereas the gravitational reading admits stable effective-crossing realisations.
\end{enumerate}

We emphasise what the argument does and does not claim. It does not claim that the data favour modified gravity over a dark-energy source: at the background level, the designer theorem renders any such evidential discrimination impossible in either direction and the perturbative information the combinations carry does not separate the readings at present precision (\S\ref{sec:breaking}).\footnote{The argument is accordingly indexed to the present evidential situation: it holds while the discriminating channels are silent, and lapses when they speak.}  Instead, its purpose is to test one possible way of deciding between the two on theoretical grounds. If the matter-sector interpretation were favoured because it offered a more straightforward theoretical realisation of the reconstructed behaviour, the phantom crossing would count against its simplest implementation. Therefore, the argument does not establish a general theoretical preference for modified gravity and no comparison of the overall simplicity of the two theory spaces follows: gravitational realisations may carry independent theoretical costs of their own. Rather, its argumentative role is to show that the phantom feature removes an otherwise immediate simplicity-based reason for favouring D3 over the gravitational reading. 

This conclusion is then deliberately circumscribed. D3 is broader than its simplest single-component realisation: more elaborate matter-sector constructions can accommodate a regular phantom crossing and their existence is not a counterexample to the argument. It is what keeps D3 in the rivalry, while demonstrating that its canonical realisation cannot provide a straightforward, simplicity-based tie-breaker.

\section{The epistemology of DESI}\label{sec:philosophy}

The case for our central claim is now complete: DESI establishes D2, thereby supports D1, and does not establish D3. What remains is to say what follows from it. One question concerns practice: how can the way in which the result is phrased as ‘evolving dark energy’ invite a slide from D2 to D3, and how should such results be phrased? Another question concerns the nature of the underdetermination that we have exhibited. We have claimed that the underdetermination is breakable, and we must provide an explanation of how this can be achieved. Also, what kind of underdetermination this is matters, because the kind determines which responses to it are available. The final question relates to the presupposition of the claim of evolving dark energy, which is a fact whose robustness itself deserves scrutiny. The subsections address these issues in turn. 

\subsection{How the reporting convention invites the slide from D2 to D3}\label{sec:theoryladen}

The argument so far shows that D2 does not by itself establish D3. It does not yet explain why the reporting convention of presenting the phenomenological CPL inference in the vocabulary of an \lq{}evolving dark-energy\rq{} may invite a stronger reading than the background data can license. The distinction is important: nothing in the CPL functional form itself selects a matter-sector interpretation. We now explain why the slide from D2 to D3 can in fact arise and how the result should be reported so as to keep the competing interpretations explicit. 

The $(w_0,w_a)$ posteriors are the quantities reproduced in plots, compared across analyses, and featured in public announcements,and they are not a theory-neutral summary of what the instrument delivered.\footnote{The point connects to a broader literature on how the format of a data product shapes what it can evidence \citep{Tal2013TheEO,tal:2016,Boyd2018}. Nothing in our argument depends on the details of those accounts.}  Some failure of neutrality is inevitable, since quantities in the inferential chain is model-mediated (\S\ref{sec:observables}),  but this does not make them ontologically committed to a dark-energy \lq{}fluid\rq{}.  The relevant failure here is of a more specific kind and enters at the level of their interpretation and reporting. Both gravitational and material readings can be assigned an effective equation of state (\S\ref{sec:fR}), but only the material one characterises the stress-energy of a genuine dark-energy source. The gravitational reading it is an effective representation of modified gravitational dynamics, measuring the mismatch between the modified law and GR. 

Accordingly, the slide does not arise from using CPL, but from reporting the fitted function specifically as the equation of state of a dark-energy source, without keeping its effective gravitational interpretation equally \textit{explicit}. This vocabulary can make the matter-sector reading appear canonical even though the same background history admits a gravitational reading.

The point does not even  require that this stronger ontological reading be intended or suggested by the collaboration. For instance, \cite{DESI2025} says that the CPL parametrisation of $w(a)$ \lq{}\lq{}does not arise directly from an underlying physical model, it is a flexible parametrization that is capable of matching the predictions for observable quantities obtained in a wide range of models that are physically motivated\rq{}\rq{} (ibid., p.21). 
The reporting convention does not by itself commit one to an ontological reading: `evolving dark energy' is standardly used for a time-dependent effective equation of state within the chosen phenomenological framework. Yet the same vocabulary also makes a stronger reading readily available, on which the fitted evolution is taken to characterise a genuine matter-sector source. Some presentations of the DESI result---including material hosted on the DESI website---adopt this stronger language explicitly \citep{BeltzMohrmann2025}.\footnote{\citet{BeltzMohrmann2025} states that \lq{}\lq{}the findings provide further indication that a cosmological constant is not the origin of cosmic acceleration, but rather that dark energy is a kind of dynamically evolving fluid that pervades all of space. With this result, a whole new era of cosmology begins\rq{}\rq{}.  \citet{Castelvecchi2025}  is more cautious, but his reference to energy density when he writes that \lq{}\lq{}the data suggest that its energy density—the amount of dark energy per cubic metre of space—is now around 10\% lower than it was 4.5 billion years ago.\rq{}\rq{}  is a clear indication of reifying language.}

The choice of reporting vocabulary is a convention, and a convention cannot by itself warrant an ontological conclusion. What it can do is make one conclusion look like the default, thereby inviting the D2-to-D3 slide without itself establishing D3. 

\subsection{What kind of underdetermination?}\label{sec:taxonomy}

Recent philosophy of cosmology has mapped the issues afflicting the dark energy case with some care. \cite{WolfDuerr2024} diagnose a \lq{}triple underdetermination\rq{} in the Dark Energy problem. 
The present case most directly relates to their first type: underdetermination of a theory by the available evidence, which is usually considered to be potentially \textit{transient}. \cite{WolfRead2026}, building on work with Ferreira \citep{WolfFerreira2023,FerreiraWolfRead2025}, and on work of \cite{Pitts2011}, analyse a different problem: a \emph{permanent} underdetermination between distinct microphysical realisations of dark energy itself. 
The authors show that distinct theories can yield empirical predictions that are arbitrarily close even when their underlying microphysical structures differ. Within the single framework of minimally coupled quintessence, physically different scalar-field potentials (such as hilltop, pseudo-Nambu--Goldstone, and others) can generate predictions in the relevant observable space, including the $(w_0,w_a)$ parameter space, that are arbitrarily close while implying radically different underlying physics and long-term cosmic futures. This permanent underdetermination arises because cosmological observations only constrain a coarse-grained effective description, leaving much of the underlying microphysics inaccessible. 

The DESI case analysed here is structurally different from this microphysical case. The issue is not which microphysical realisation underlies a given dark-energy phenomenology, but whether that phenomenology should be attributed to a genuine matter-sector dark energy source within GR at all. The rivals are not two realisations within a common quintessence framework, but rather two accounts of the same expansion history: an evolving matter-sector source within GR and modified gravitational dynamics. The underdetermination therefore arises at the level of the explanatory framework rather than at the level of the underlying microphysics, though each framework is instantiated by fully specified models: a \wCDM\ model on one side, a designer $f(R)$ model reproducing the same $H(z)$ on the other.\footnote{This is, moreover, precisely a rivalry that lies outside the scope of the common-core resolution developed by \cite{WolfRead2026}: their argument applies only within the restricted canonical, minimally coupled quintessence subspace considered there, leaving open the underdetermination between that class and modified-gravity proposals. \cite{WolfDuerr2024} note the same rivalry in a footnote of their own, and suggest that checks of non-minimal coupling might break the degeneracy. We show that at the level of background observables it cannot be broken at all, and that the checks they point to are now under way.}

This shift in the relata forfeits an immunity that the microphysical case enjoys. Permanent underdetermination is immune by construction to the charge that the rivals are merely notational variants of one another, since they are empirically inequivalent \citep{WolfRead2026}. A rivalry claimed to be exact within a channel is not, and here the charge takes a concrete form. $f(R)$ gravity admits an \textit{Einstein frame} presentation in which it is GR plus a scalar field minimally coupled to gravity \citep{Faraoni1999}. Is the `gravitational' reading, then, not itself a dark-energy matter model, so that D3 is vindicated after all? Answering requires distinguishing between: (i) GR plus a component interacting with the standard sectors \emph{only gravitationally}, whose equation of state is the reconstructed $w(z)$, and (ii) the designer $f(R)$ construction with its two presentations, Jordan and Einstein. The Einstein frame presentation of (ii) is \textit{not} (i). Its scalar, while minimally coupled to gravity, is directly coupled to matter, so that $f(R)$ is equivalent to \emph{coupled}, not minimal, quintessence \citep{Pettorino2008}. The operative contrast, here and in D3, is therefore drawn where a standard demarcation in the physics literature draws it \citep{Joyce2016}: between theories that are GR plus purely gravitationally coupled components, and theories that are not. Regardless of which conformal representation is used, the theory fails to instantiate D3 as defined here. In the Jordan frame matter is minimally coupled but the gravitational dynamics are non-Einsteinian. In the Einstein frame the gravitational action takes Einstein--Hilbert form but the scalar is conformally coupled to matter. No conformal representation therefore turns the designer $f(R)$ model into GR plus a dark-energy degree of freedom coupled to the standard sectors only gravitationally. The conformal transformation redistributes the scalar interaction between the gravitational and matter descriptions, it does not remove it. The rival designs therefore remain on the modified-gravity side of the contrast.\footnote{\cite{Norton2008} sets out two deflationary challenges. The first is general: wherever observational equivalence can be demonstrated tractably, the theories are close enough in structure that we cannot rule out the possibility that they are variant formulations of one theory. The second challenge is more specific and relates to artificially generated rivals, whose construction may represent an unnecessary simplification that is not equally supported by the evidence. This challenge is answered by the division of labour of \S\ref{sec:viability}, since the gravitational alternative is supported by independently developed theories rather than by the designer family. The first challenge does not undermine the conclusion but restates it: the two alternatives are equivalent on the background channel, but not beyond it. In other words, background observables do not fix the difference between them.}

The second difference from the Wolf--Read case concerns the character of the equivalence. In their case the rival models are empirically inequivalent in principle and become indistinguishable only because their predictions can be brought arbitrarily close within the accessible regime. The designer theorem instead gives exact coincidence: for any admissible expansion history $H(z)$ there are designer $f(R)$ models reproducing the background observables \textit{exactly}. What is exact is not the empirical equivalence of the two frameworks as a whole, but their coincidence on the observables the background channel delivers, and that coincidence follows from the structure of the field equations rather than from finite observational precision. Therefore, the two frameworks are not empirically equivalent simpliciter. They agree only on the observables determined solely by the homogeneous expansion history. Perturbations, lensing observables, growth measurements, standard sirens, and laboratory tests of gravity provide, at least in principle, ways of distinguishing the rivals, though only relative to the physically motivated model classes actually in play.\footnote{A maximally flexible fluid endowed with free anisotropic stress could in principle evade even these probes.}
Crucially, whether a given designer reconstruction is free of pathology has to be checked case by case (\S\ref{sec:designer}), and the non-minimally coupled theories that support the gravitational reading match the data at current precision rather than reproducing the best fit exactly (\S\ref{sec:viability}). The rivals are therefore not empirically closer than Wolf and Read's. What makes them indiscriminable is not the nearness of distinct predictions but a channel that is provably blind to the distinction.

To conclude, there are two taxonomies at play here, and the case falls differently under each one. For \cite{WolfDuerr2024} it is evidential underdetermination, their first kind. 
In the terms of \cite{WolfRead2026} it is not permanent underdetermination, since the indiscriminability does not come from the impossibility to tighten the bounds towards zero under finite precision. 
Crucially, it is neither \textit{strong} underdetermination \emph{simpliciter}, since the two frameworks are not empirically equivalent across all observables.\footnote{Strong underdetermination is a key concept in classical philosophy of science, associated primarily with the work of \cite{Duhem1954,Quine1975}. In short, strong underdetermination is the perfect identity of observational outcomes, whereas permanent underdetermination is the practical indistinguishability that arises from the intrinsic limitations of our ability to measure.} However, when testricted to the background channel, it has the structure of strong underdetermination: the empirical substructures fixed by $H(z)$ coincide exactly, meaning no improvement in the precision of the distance measurements that carry the DESI result can discriminate between the readings. Therefore, the case may be understood as one of strong underdetermination confined to a single evidential channel and breakable outside it, relative to the model classes in play. 

This yields an immediate assessment of the stronger ontological reading of the DESI result. If ‘evidence for dynamical dark energy’ is read as asserting D3, that reading goes beyond what D2 establishes. 
D2 is compatible with at least two distinct explanatory frameworks, one invoking an evolving dark-energy component of the kind specified by D3 and the other invoking modified gravitational dynamics. The problem is therefore not merely that the evidence remains incomplete, but that the background data themselves do not discriminate between the competing explanations. 
Where Wolf and Read's quintessence case shows that we may \textit{never} know which microphysical realisation of dark energy is correct, the present case shows that no measurement of the background expansion history, \textit{at any precision}, could establish that the DESI-favoured departure from the \LCDM\ expansion history is produced by a genuine dynamical dark-energy source in the matter-sector. 

The classification also fixes what can be done about it. The responses available to permanent underdetermination are non-empirical, turning on discrimination, common cores, or overarching frameworks \citep{LeBihan2018-LEBTL,WolfRead2026}. In \cite{WolfRead2026}'s sense, discrimination seeks to privilege one of the underdetermined alternatives by appeal to considerations such as theoretical virtues, explanatory power, or the absence of pathological structure. The phantom-crossing argument of \S\ref{sec:asymmetry} tests a candidate strategy of this kind, namely whether theoretical straightforwardness can break the tie in favour of the matter-sector reading. Ultimately, however, it cannot, as the phantom feature removes the immediate reason for privilege, yet leaves the inter-framework rivalry unresolved. Because the present underdetermination is confined to a single channel, an empirical response is available instead. Section~\ref{sec:breaking} sets out what it consists in.

\subsection{Breaking the underdetermination: perturbations, sirens, and the laboratory}
\label{sec:breaking}

That the underdetermination is breakable is not a vague promissory note: the relevant observables are known and, in some cases, already measured. The key point is simple. The designer theorem (\S\ref{sec:designer}) guarantees agreement only at the level of the homogeneous FLRW background. It does \emph{not} ensure that a designer $f(R)$ and a GR model with a matter-sector dark energy source respond in the same way to inhomogeneities. Two cosmologies can therefore agree perfectly on the homogeneous expansion while differing in how matter clusters, how light propagates through large-scale structure, and how gravitational waves travel across cosmological distances.

This distinction is captured by \textit{cosmological perturbation theory} \citep{Bertschinger2008}. 
The same expansion history need not imply the same gravitational response to inhomogeneities: theories that agree on $H(z)$ can predict different rates of structure growth and different lensing effects. 
Modified-gravity analyses commonly parametrise these departures through parameters such as $\mu$, which characterises the gravitational response to matter perturbations and $\Sigma$, which characterises the lensing response.\footnote{More precisely, $\mu(a,k)
\equiv
\frac{G_{\mathrm{eff}}(a,k)}{G}$, where $a$ is the scale factor, $k$ the comoving wavenumber of the perturbation, $G$ Newton's constant, and $G_{\rm eff}(a,k)$ the effective gravitational coupling governing the response of the metric to matter perturbations. The lensing parameter $\Sigma$ is related to $\mu$ and the gravitational slip parameter $\eta$ by $\Sigma(a,k)\equiv\frac{\mu(a,k)}{2}\left[1+\eta(a,k)\right]$, with $\eta(a,k)\equiv\frac{\Phi}{\Psi}$. Here $(\Phi,\Psi)$ are the two scalar gravitational potentials appearing in the perturbed metric in Newtonian gauge, $ds^2=-(1+2\Psi)dt^2+a^2(1-2\Phi)d\mathbf{x}^2$. We refer the reader to e.g. \cite{DeFelice2010,AmendolaKuntzSapone} for technical details.}
In GR with a minimally coupled and sufficiently smooth dark-energy source---canonical quintessence being the simplest example---both retain their GR values ($\mu=\Sigma=1$), whereas modified-gravity models can alter them without altering the background history. Redshift-space distortions and galaxy clustering probe the former type of information, while weak and CMB lensing probe the latter.

These data therefore probe a dimension along which background-degenerate rivals can differ. The DESI collaboration has already carried out a dedicated analysis along this dimension. At current precision, the modified-gravity parameters tested are consistent with their GR values, constraining but not eliminating the gravitational alternatives relevant here \citep{DESIMG2024}.\footnote{For related analyses, see \citet{Chudaykin2024,Ye2025}, which explore modified gravity as a viable explanation of the DESI preference.} The gravitational candidates of \S\ref{sec:viability} have moreover been fitted to data combinations containing perturbation-sensitive information and remain competitive. Perturbative observables can thus provide discriminating evidence that is absent from the background channel.

There is, however, an important limit to this conclusion.  A sufficiently general GR stress-energy, endowed with suitable perturbations and anisotropic stress, can mimic modified-gravity signatures even at the perturbative level \citep{KunzSapone2007}. Perturbative observables therefore do not provide a universal discriminator between D3 and modified gravity \textit{simpliciter}. Their discriminating power depends on the model classes specified: they can distinguish specified matter-sector models from specified gravitational rivals when those models make different predictions for growth or lensing.  This qualification is precisely what the taxonomy of \S\ref{sec:taxonomy} leads us to expect. The underdetermination is breakable, but no single perturbative observable is guaranteed to break it across an unrestricted space of theories. This is why independent channels are epistemically valuable: they test features of the rival frameworks that are not exhausted by their effective cosmological stress-energy description.

A second route to discrimination comes from \textit{gravitational-wave cosmology}. Standard sirens---the gravitational-wave analogue of standard candles---infer an absolute luminosity distance from the gravitational-wave signal, without relying on the cosmic distance ladder. In theories with an evolving effective gravitational coupling, gravitational waves propagate differently from electromagnetic signals. The result is that the gravitational-wave luminosity distance $d_L^{\mathrm{GW}}(z)$ can differ from the electromagnetic luminosity distance $d_L^{\mathrm{EM}}(z)$, whereas any minimally coupled dark-energy model within GR predicts equality \citep{Belgacem2018}. Standard sirens therefore probe information about gravitational propagation that is not fixed by the background history $H(z)$.

A third route takes the comparison outside cosmological phenomenology altogether, through Solar-System and laboratory tests. Fifth-force searches and screening tests probe the scalar degree of freedom directly, at densities and curvatures where the background degeneracy offers no shelter. This channel is already empirically active: local tests place severe restrictions on familiar screened $f(R)$ models such as Hu--Sawicki (\S\ref{sec:viability}), and analogous constraints bear, model by model, on the non-minimally coupled rivals. Discrimination along this dimension has already begun, but is not concluded: whether the non-minimally coupled rivals admit an acceptable screening mechanism is at present an open question of model-building, and the answer can come from this independent channel rather than from the expansion history. 

The upshot is not that D3 is empirically inaccessible. It is that background observables alone cannot decide it. Moving beyond the background opens several independent dimensions of empirical comparison: structure growth and lensing test the gravitational response to inhomogeneities; standard sirens can test gravitational-wave propagation; and local experiments constrain additional interactions and their screening. 
The corresponding reporting norm is therefore straightforward: a result should be reported at the strongest inferential level actually warranted by the observables and assumptions used, while any stronger physical interpretation should be identified explicitly as an interpretation rather than presented as part of the result itself. In fact, stronger physical language becomes warranted only insofar as additional observables discriminate between the relevant matter-sector and gravitational-sector alternatives. Constraints on the presented perturbative quantities are precisely the kinds of additional evidence that can perform this discriminating work \citep{DESIMG2024}. 

\subsection{How robust is the explanandum?}\label{sec:robustness}

So far, our argument has granted the D1-level claim that the late-time expansion history genuinely departs from \LCDM. 
We then focused on the question of what, if such a departure exists, would explain it. 
However, before drawing conclusions, it is worth considering whether there is, in fact, a robust departure from the \LCDM background, since this sets the strength of the explanandum to which any interpretation must respond.

At present, the empirical situation is promising but not conclusive \citep{Wang2025}. DESI BAO measurements alone remain compatible with \LCDM and the preference for a time-dependent equation of state becomes appreciable only when DESI is combined with external information, especially CMB and SN~Ia data. The strength of this preference depends on which supernova compilation is used \citep{DESI2025,CortesLiddle2025}. This dependence should not be overstated. The DESI supporting analysis reveals a consistent qualitative trend across various alternative parametrisations and non-parametric reconstructions, suggesting that the effect is not simply an artefact of the CPL ansatz \citep{Lodha2025}. However, the dependence should not be ignored either. Reanalyses of the supernova data have dependencies on the treatment of the low-redshift sample, as well as on calibration and modelling choices \citep{Efstathiou2025,Vincenzi2025,Capozziello2026}. Furthermore, theory-informed priors can substantially reduce the apparent preference for evolving dark energy \citep{Toomey2026}. While independent DES results point in the same phenomenological direction, they currently provide corroboration rather than decisive replication \citep{DESY62026}.

The appropriate conclusion is therefore neither that the anomaly is spurious nor that it is already secure. Its empirical status remains provisional. While this is important, it only has a limited impact on the argument of this paper. If the departure from LCDM were to be disproved, there would simply be no anomalous expansion history requiring the explanations considered here. Conversely, if the departure from LCDM were to become overwhelming and independently established, the central underdetermination would remain. Increasing the evidential support for a non-\LCDM\ background does not by itself determine what physical ontology or gravitational dynamics produces that background.

This distinction is crucial to the inferential structure developed above. The robustness of D1 concerns whether there is an explanandum and it is currently under investigation, while the D2$\to$D3 problem concerns what may legitimately be inferred about its explanation. Therefore, uncertainty about the anomaly, or its future confirmation, does not affect the central result: background evidence can support a departure from \LCDM\ without establishing that the departure is due to genuine matter-sector dark-energy degrees of freedom within GR.

\section{Conclusions}
\label{sec:conclusion}

The DESI results are a major achievement of observational cosmology. They provide increasingly precise information about the late-time expansion history of the universe and have placed renewed pressure on the standard \LCDM\ model. The central claim of this paper has been that the inferential significance of these results must nevertheless be stated at the appropriate level. DESI-based analyses establish a framework-internal parametric preference, D2, which in turn supports a genuine departure from the \LCDM\ background, D1. Neither result, however, establishes D3: that the departure is produced by one or more genuine dynamical dark-energy degrees of freedom in the matter-sector, within GR and coupled to the standard sectors only gravitationally. This conclusion does not depend on the present statistical security of the anomaly. 

The reason is not merely that current observations are insufficiently precise. The designer construction in $f(R)$ gravity shows that, for the relevant class of expansion histories, modified gravitational dynamics can reproduce exactly the same homogeneous background as a GR model in which the departure is attributed to a dark-energy source. Background observables therefore cannot identify which of these explanatory frameworks is responsible for the observed history, however accurately that history is measured. 

This also locates the DESI case within the broader philosophical literature on underdetermination in cosmology. The rivalry considered here is not between different microphysical realisations of dark energy within a common framework. It concerns whether the same cosmological phenomenology should be attributed to a matter-sector source within GR or to a modification of gravitational dynamics. The resulting underdetermination is accordingly neither an ordinary problem of finite precision within this observational channel nor an empirical equivalence between the frameworks as a whole. It is exact at the level of the background and breakable in principle outside it relative to the model classes in play.

We also showed that the gravitational reading does not rest on the designer family alone: independently developed non-minimally coupled theories realise the relevant phenomenology, admit stable effective phantom-crossing behaviour and fit the same data combinations comparably to the matter-sector template, although their observational viability---in particular their compatibility with local tests and the availability of satisfactory screening---remains unsettled. 
The phantom crossing then provides a stress test of a possible theoretical preference for the matter-sector reading. If theoretical straightforwardness is invoked to privilege that reading, the crossing then would place substantial pressure precisely on its simplest single-component, minimally coupled realisation, which cannot produce it without pathology, whereas the gravitational-sector can. This does not favour modified gravity over D3 as a whole, since more elaborate matter-sector models can also realise such behaviour. It does remove one possible reason for treating the simplest matter-sector realisation as the theoretically privileged interpretation of the background result. 

The epistemological lesson extends beyond this particular model comparison. Cosmological observations are necessarily reported through theoretical and phenomenological representations. The CPL parameters $(w_0,w_a)$ are legitimate and useful tools for compressing information about the expansion history, and the parametrisation itself does not privilege a matter-sector over a gravitational interpretation. A parametrisation does not acquire additional evidential force simply because a \lq{}substance\rq{} vocabulary is attached to it. Therefore, the appropriate reporting norm we outline is to clearly distinguish between the phenomenological quantities constrained by the data and stronger claims about the physical mechanism that produces them.

The relevant empirical question is consequently not only whether the late-time expansion differs from \LCDM, but what produces any such departure. Growth and lensing, standard sirens, and local tests of gravity supply the channels along which that question can be answered. They are already in play and at present precision they constrain the gravitational alternatives without eliminating them. Their discriminating power is moreover relative to the model classes specified, since no single observable separates the readings across an unrestricted space of theories. Stronger claims about the physical origin of the departure should therefore track the evidence these channels supply. 

The present situation is accordingly best described without either deflating or giving ontological interpretations to the DESI result. There is a serious, but still provisional, indication that the late-time expansion history may depart from that of \LCDM. Identifying its cause constitutes a further empirical challenge. DESI and future experiments may reveal new cosmic dynamics. What remains open is its physical origin.

\section*{Acknowledgments}
I thank Giovanni Montani for useful discussions.

\bibliography{biblio}

\end{document}